\documentclass[11pt,document,nofootinbib,superscriptaddress,onecolumn,preprintnumbers,balancelastpage]{article}

\pdfoutput=1

\usepackage{jheppub}

\usepackage{subfig}
\usepackage[countmax]{subfloat}
\usepackage{framed}
\usepackage{mdframed}
\usepackage{multirow}
\usepackage{slashed}
\usepackage{booktabs} 

\usepackage{array}

\makeatletter  
\newcommand{\thickhline}{%
    \noalign {\ifnum 0=`}\fi \hrule height 1pt
    \futurelet \reserved@a \@xhline
}

\newcolumntype{"}{@{\hskip\tabcolsep\vrule width 1pt\hskip\tabcolsep}}

\makeatother
\def\beq{\begin{eqnarray}}
\def\eeq{\end{eqnarray}}
\def\bea{\begin{eqnarray}}
\def\eea{\end{eqnarray}}
\usepackage{latexsym}
\usepackage{amssymb}
\usepackage{epsfig,amsmath,graphics}
\usepackage{listings}
\usepackage{color}
\usepackage{dcolumn}
\usepackage{slashed}
\usepackage{cancel}
\usepackage{comment,latexsym} 
\usepackage{bm}
\usepackage{verbatim}
\usepackage{tabularx}
\usepackage{dcolumn}
\definecolor{Mahogany}{rgb}{0.62,0.24,0.15}
\definecolor{colorLink}{rgb}{0.7,0,0}
\definecolor{colorCite}{rgb}{0,.7,0}
\definecolor{colorURL}{rgb}{0,0,0.7}
\usepackage{xspace}
\usepackage{multirow}
\usepackage{titlesec}
\usepackage[bbgreekl]{mathbbol}
\usepackage{bm}
\usepackage{float}
\usepackage{placeins}
\usepackage{afterpage}
\usepackage{footmisc}
\usepackage{breakcites}

\usepackage{empheq}
\usepackage[most]{tcolorbox}
 \tcbset{myformula/.style={colback=white!40,colframe=black,
every box/.style={highlight math style={colback=LightBlue!50!white,colframe=Navy}}
}}

\usepackage[toc,page]{appendix}

\usepackage{etoolbox}
\appto\appendix{\addtocontents{toc}{\protect\setcounter{tocdepth}{1}}}

\usepackage{scalerel}

\usepackage{arydshln}

\allowdisplaybreaks

\expandafter\def\expandafter\normalsize\expandafter{%
    \normalsize
    \setlength\abovedisplayskip{8pt}
    \setlength\belowdisplayskip{8pt}
    \setlength\abovedisplayshortskip{8pt}
    \setlength\belowdisplayshortskip{8pt}
}

\newcommand{\be}{\begin{equation}}
\newcommand{\ee}{\end{equation}}

\newcommand{\gsim}{\lower.7ex\hbox{$\;\stackrel{\textstyle>}{\sim}\;$}}
\newcommand{\lsim}{\lower.7ex\hbox{$\;\stackrel{\textstyle<}{\sim}\;$}}

\newcommand{\nocontentsline}[3]{}
\newcommand{\tocless}[2]{\bgroup\let\addcontentsline=\nocontentsline#1{#2}\egroup}

\newcommand{\Msun}{M_\odot}
\newcommand{\OM}{\Omega_{\rm M}}
\newcommand{\OCDM}{\Omega_{\rm CDM}}
\newcommand{\fPBH}{f_{\rm PBH}}

\title{
Exploring the Early Universe with Gaia and THEIA
}

\author[a]{Juan Garcia-Bellido,}
\author[b,c]{Hitoshi Murayama}
\author[d]{and Graham White}

\affiliation[a]{\footnotesize Instituto de F\'isica Te\'orica UAM/CSIC, Universidad Aut\'onoma de Madrid, Cantoblanco, 28049 Madrid, Spain}
\affiliation[b]{\footnotesize Berkeley Center for Theoretical Physics, University of California, Berkeley, CA 94720, USA}
\affiliation[c]{\footnotesize Theory Group, Lawrence Berkeley National Laboratory, Berkeley, CA 94720, USA}
\affiliation[d]{\footnotesize IPMU, University of Tokyo, 5-1-5 Kashiwanoha, Kashiwa, Chiba, 277-8583, Japan}

\emailAdd{juan.garciabellido@uam.es}
\emailAdd{hitoshi@berkeley.edu}
\emailAdd{graham.white@ipmu.jp}

\abstract{It has recently been pointed out that Gaia is capable of detecting a stochastic gravitational wave background in the  sensitivity band between the frequency of pulsar timing arrays and LISA. We argue that Gaia and THEIA has great potential for early universe cosmology, since such a frequency range is ideal for probing phase transitions in asymmetric dark matter, SIMP and the cosmological QCD transition. Furthermore, there is the potential for detecting primordial black holes in the solar mass range produced during such an early universe transition and distinguish them from those expected from the QCD epoch. Finally, we discuss the potential for Gaia and THEIA to probe topological defects and the ability of Gaia to potentially shed light on the recent NANOGrav results.}

\begin{document} 

\maketitle

\section{Introduction}
It has recently been shown that large surveys of stars such as Gaia~\cite{Brown:2018dum} and the proposed upgrade, THEIA (Telescope for Habitable Exoplanets and Interstellar/Intergalactic Astronomy)~\cite{10.3389/fspas.2018.00011}, can be powerful probes of gravitational waves (GW)~\cite{Moore:2017ity}. GWs affect the apparent position of a star, and the multiple subsequent measurements of the same star can be used to turn Gaia into a GW observatory~\cite{Moore:2017ity,Mihaylov:2018uqm,Mihaylov:2019lft}. The dimensionless strain sensitivity of Gaia is expected to be a constant~\cite{Moore:2017ity} and scale inversely with the lifetime of the experiment~\cite{Book:2010pf}. This is in contrast to pulsar timing arrays, whose strain sensitivity scales as the inverse square root of the observation time. The scaling with the mission lifetime comes from the fact that Gaia monitors the position of $N$ sources in the sky with angular resolution $\Delta\theta$ over a time $T$. For a single source, one could detect an angular
velocity (proper motion) of order $\Delta\theta/T$, and for $N$
sources, a {\em correlated} angular velocity of order $\Delta\theta/(T\sqrt{N})$ should be detectable. It follows~\cite{Book:2010pf} that one should obtain an upper limit 
\begin{equation}
    \Omega_{\rm gw}(f\sim 1/T) \lsim \frac{\Delta\theta^2}{NT^2H_0^2}
\end{equation}
on the GW energy density from a SGWB. Moreover, the analysis of Ref.~\cite{Moore:2017ity} seems to indicate that the cadence of the astrometric survey does not change the strain sensitivity appreciably, at least not more than an order of magnitude.
\par 
The constant strain sensitivity makes astrometry a powerful tool for filling the sensitivity gap between nanohertz frequencies probed at pulsar timing arrays~\cite{Lentati:2015qwp,Arzoumanian:2020vkk}, and millihertz frequencies probed by the LISA mission~\cite{Bartolo:2016ami}. Furthermore, the efficient scaling with the mission time makes upgrades to Gaia competitive with even the square kilometer array at probing the nanohertz range. In this paper we discuss the cosmological opportunities of using Gaia and THEIA as GW observatories. In particular, various dark matter models based on the Strongly-Interacting Massive Particles (SIMP) paradigm are expected to feature a confining transition that could leave a observable background gravitational wave spectrum that typically peaks between the sensitivity ranges of LISA and pulsar timing arrays, precisely the frequency range where astrometry can contribute. Asymmetric dark matter is also expected to include physics at a similar scale. Further we explore the potential of primordial black hole detection with an emphasis of primordial black hole production during such a confining transition. We also discuss the potential to observe gravitational wave backgrounds generated from topological defects such as domain walls and strings. Finally, motivated by the recent hint of a possible SGWB signal at NANOGrav \cite{Arzoumanian:2020vkk}, we discuss the timescale for Gaia to give complementary information on the tentative signal. \par

The structure of this paper is as follows. Section \ref{sec:curves} we give the sensitivity curves for Gaia and its upgrades, we then discuss the potential to observe cosmological phase transitions indirectly with astrometry in Section \ref{sec:PT}. Note that in that section we discuss specific models that can be detected, focusing on dark sectors. Next, in section \ref{sec:defects}, we discuss cosmological defects and the potential reach of astrometry as well as a discussion of the recent NANOGrav results. In section \ref{sec:pbh} we discuss the effect that phase transitions have on the Primordial Black Hole (PBH) mass spectrum as a complementary cosmological probe before concluding in section \ref{sec:conclusions}. We describe in the Appendices the low-energy effective models associated with confinement transitions.

\section{Sensitivity curves for Gaia and THEIA}\label{sec:curves}
\begin{figure}
    \centering
    \includegraphics[width=\textwidth]{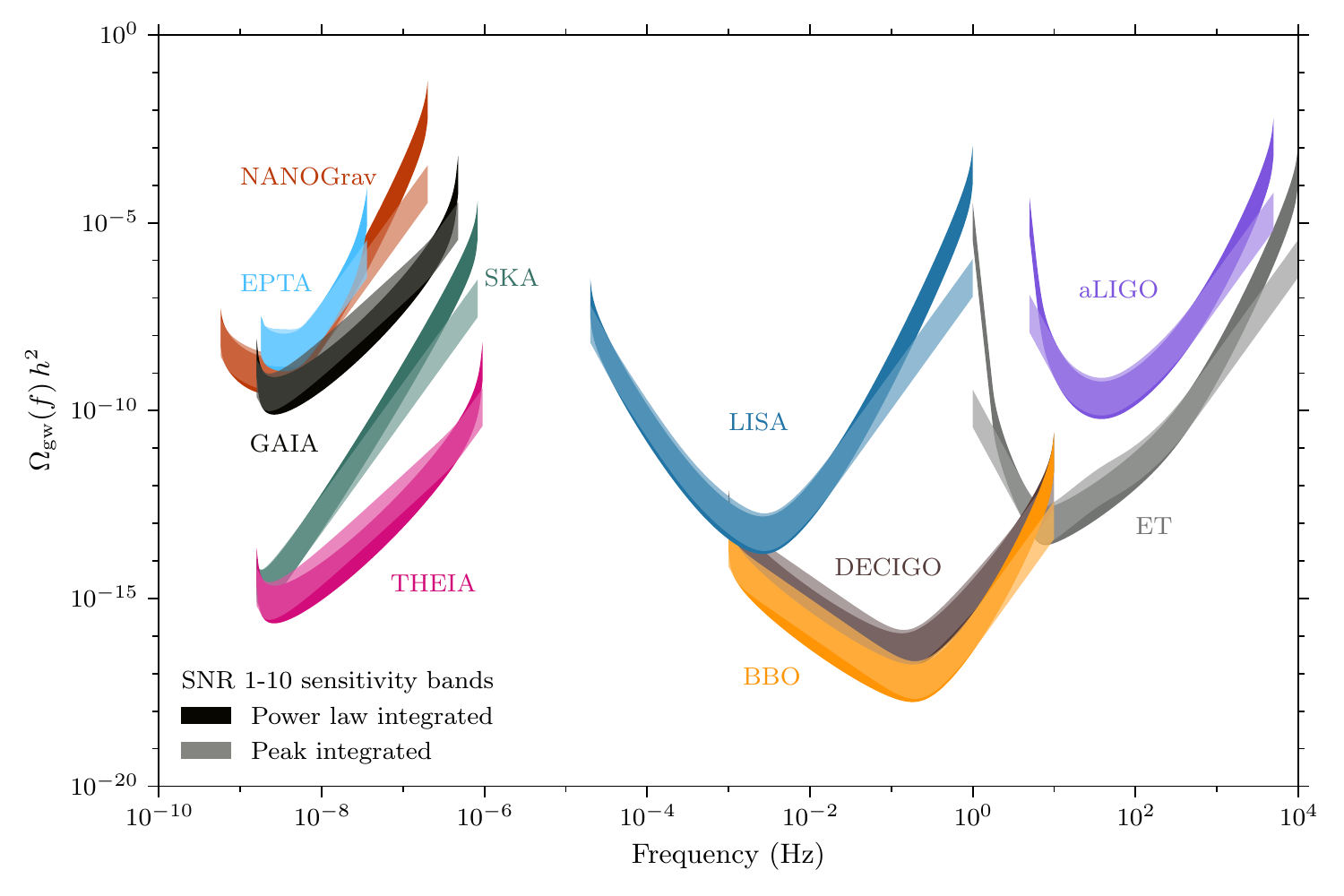}
    \caption{Power law and peak integrated sensitivity curves for pulsar timing arrays \cite{vanHaasteren:2011ni,Janssen:2014dka,Lazio:2018ukb} against astrometry methods such as Gaia and THEIA. The bands correspond to a signal to noise ratio between 1 and 10, the lighter bands are power law integrated sensitivity curves while the darker bands correspond to the peak integrated sensitivity bands. Also included are space based missions, Lisa \cite{Caprini:2019egz}, BBO \cite{Yagi:2011wg} and Decigo \cite{Kawamura:2006up,Kawamura:2020pcg} where for the latter we have used the ultimate specifications. Additionally, we included high frequency detectors in aLIGO \cite{TheLIGOScientific:2014jea} and the Einstein telecope \cite{Punturo:2010zz}. Note that THEIA has sensitivity in the range beween pulsar timing arrays and LISA. By extending the cadence of THEIA down to days one could fill the gap between PTA and LISA frequencies.}
    \label{fig:sensitivity}
\end{figure}

Experimental designs typically specify a strain sensitivity, whereas cosmologists are typically interested in the gravitational wave abundance.
The strain sensitivity, $h_{\rm gw}$ can be conveniently converted to the abundance, $\Omega _{\rm GW}$ via the relation
\begin{equation}
    \Omega_{\rm gw}(f)\,h^2 = \frac{2 \pi^2}{3 H_0^2} f^2 h_{\rm gw}^2(f) h^2
\end{equation}
where $h = 0.67$, i.e. $H_0=67$ (km/s)/Mpc, is the value of Hubble observed today and $f$ is the frequency. For Gaia we can take the constant sensitivity derived in ref. \cite{Moore:2017ity}
 \begin{equation} h_{\rm GW}=10^{-14} \left( \frac{5 \ {\rm years}}{T_M} \right), \ \forall \ f>1/T_M 
\end{equation} 
where $T_M$ is the mission lifetime, which we will generously take to be $~20$ years. For THEIA, we estimate that one will observe a hundred times as many stars with an angular velocity resolution that is superior by a factor of 60. This results in a strain sensitivity of 
\begin{equation} h_{\rm GW}=1.6\times 10^{-16} \left( \frac{5 \ {\rm years}}{T_M} \right), \ \forall \ f>1/T_M \ . \end{equation}
When searching for a specific signal, the sensitivity of Gaia can be dramatically improved by comparing the signal for multiple frequencies.  We present the peak integrated sensitivity curve~\cite{Schmitz:2020syl} relevant for phase transitions and the power law sensitivity curve~\cite{Thrane:2013oya} relevant for strings in Fig.~\ref{fig:sensitivity}. In both cases the integrated sensitiviity curve can be obtained by
\begin{eqnarray}
  {\cal N} (f)  &=&\left(3 H_0^2 \int _{f_{\rm min}=1/T} ^{f_{\rm max}} \left(\frac{f^\prime}{f}\right)^3 \left( \frac{7}{4+3 \left(4+3 \left( \frac{f^\prime}{f} \right)^2 \right)} \right)^{7/2} \frac{df^\prime}{2 \pi ^2 (f^\prime )^3 h_{\rm GW}^2 h ^2} \right)^{-1} \\
        {\cal N} (f)  &=& {\rm Max}(n) \left[ \left( \frac{f}{f_{\rm min}} \right)^{n}  \left(3H_0^2\int _{f_{\rm min}=1/T} ^{f_{\rm max}}  \frac{(f^\prime /f_{\rm min})^n }{2 \pi ^2 (f^\prime )^3 h_{\rm GW}^2 h ^2} df^\prime \right)^{-1} \right] \,,
\end{eqnarray} 
for a broken and an envelope of power laws respectively. The sensitivity we show in Fig.~\ref{fig:sensitivity} alongside pulsar timing arrays as well as other proposed experiments including Lisa and aLIGO. THEIA is almost unmatched in its potential sensitivity and both astrometry experiments have reach in between that seen by pulsar timing arrays and LISA.

\section{Phase transitions}\label{sec:PT}

For any given phase transition, the dominant source is widely believed to be the acoustic source \cite{Hindmarsh:2013xza,Hindmarsh:2015qta,Hindmarsh:2017gnf}.  The spectral form is model independent with the peak amplitude and frequency depending just on four thermal parameters - the bubble wall velocity, $v_w$, the inverse timescale of the transition, $\beta$, the change in the trace anomaly normalized by the radiation density, $\alpha$, and the temperature at which percolation occurs, $T_p$. The peak amplitude depends on the thermal parameters as follows \cite{Hindmarsh:2013xza,Hindmarsh:2015qta,Hindmarsh:2017gnf,Guo:2020grp}\footnote{Note that the method we use in this paper is common, but has some non-trivial uncertainties, as it neglects vorticity and reheating effects \cite{Cutting:2019zws}, assumes a bag model \cite{Giese:2020znk,Giese:2020rtr} and is not always an accurate estimate of the nucleation temperature \cite{Guo:2021qcq}}
\begin{eqnarray}
h^2 \Omega_{\text{GW}} =  8.5 \times 10^{-6} \left(\frac{100}{g_{s}(T_e)}\right)^{1/3} \Gamma^2 \bar{U}_f^4  
\left[
\frac{H_s}{\beta(v_w)}
\right]
v_w 
\times
\Upsilon ,
\end{eqnarray}
where $\Gamma \sim 4/3$ is the adiabatic index and ${\Upsilon}$ 
is the suppression factor arising from the finite lifetime, $\tau_{\text{sh}}$, of the sound waves, given by
\begin{equation}
\Upsilon = 1 - \frac{1}{\sqrt{1 + 2 \tau_{\text{sh} H_s}}} .
\end{equation}

\subsection{Motivating first order transitions detectable with astrometry}
The range of frequencies that Gaia is sensitive to is ideal for a phase transition occurring between an MeV and a GeV. This is precisely the scale we would expect a confinement transition to occur in either QCD, a hidden sector involving a strongly interacting massive particle (SIMP), or asymmetric dark matter model. In fact, the preferred confinement scale for strongly interacting massive particle (SIMP) dark matter is precisely in this range \cite{Hochberg:2014dra}. We will focus mostly on hidden sector models here, but let us begin with the QCD transition, which can also in principle be a strong first order transition and can give us insight to what sort of phase transitions can lead to a gravitational wave signature.

\par 
In the standard model, the cosmological QCD transition is unfortunately not expected to be strongly first order \cite{Bazavov:2011nk,Gupta:2011wh}. However, in principle, extensions to the standard model or unusual conditions in the early Universe could modify this. Confinement transitions are not as easy to model as transitions involving fundamental scalars, where effective potential methods provide a powerful and intuitive guide and are known to describe low scale QCD reasonably well \cite{GellMann:1960np,Nambu:1961tp,Nambu:1961fr,Fukushima:2003fw}.\footnote{Note that transitions concerning fundamental scalars involve their own set of theoretical problems that can create large uncertainties in perturbative methods \cite{Croon:2020cgk}.} In the absence of lattice calculations, it is a natural to consider an effective scalar theory and model confinement transitions in an analogous manner. From such an effective theory, one can make a qualitative check of that nature of the transition by searching for stable infrared fixed points \cite{Pisarski:1983ms}. 
Between lattice calculations and effective methods of the condensate one can derive three scenarios that can make a cosmological confinement transition first order
\begin{itemize}
    \item Having at least three quarks that are effectively massless compared to the confinement scale. In QCD, this can occur if $u$, $d$, $s$ quarks
    are made lighter in the early Universe \cite{Davoudiasl:2019ugw}.
    \item Having a pure Yang--Mills transition where quarks are either absent or heavy.
    \item The baryon chemical potential is large in the early Universe. This can be achieved through a large lepton asymmetry \cite{Schwarz:2009ii} which can be made compatible with BBN bounds \cite{Barenboim:2016shh}. In principle a confinement transition in a SIMP model can be similarly influenced by a large asymmetry in the early Universe.
\end{itemize}

\par

Although the calculations we present here will suffer from theoretical uncertainty, results do seem to agree with the intuition that the larger the change in relativistic degrees of freedom, the stronger the transition. For example, in the case where glueballs dominate the confining transition, the strength of the transition grows with the number of colors. Similarly, in the case where quark condensates dominate the free energy, prior work has shown some evidence that strength of the transition grows with the number of flavours \cite{Croon:2019iuh}. We will also consider the implications of a transition of a fundamental scalar in an asymmetric dark matter model introduced in ref. \cite{Croon:2019rqu}.  \par 

\subsubsection{Gravitational waves from chiral symmetry breaking}

To estimate the gravitational wave signal in a chiral symmetry breaking transition, we follow the strategy of recent work that relies on low energy effective theories such as the linear sigma model (LSM) and the (p)NJL model \cite{Helmboldt:2019pan}. The NJL has the advantage of being sensitive to the number of colours in the theory whereas the LSM is sensitive only to the number of quark flavours. The pNJL model improves on the NJL model yet again by including a potential for Wilson loops whose coefficients can be derived from lattice \cite{Roessner:2006xn}. There is to date no lattice calculation to compare the predictions for the gravitational wave signal produced in either model, we therefore simply calculate benchmarks in each model with the expectation that there can be future improvement in the predictions of gravitational waves produced in SIMP models. 
\par 
For the linear sigma model, a large gravitational wave signal is predicted in the case where there is a large hierarchy between the axion and pion mass \cite{Croon:2019iuh}. In the case of the (p)NJL model, there is a critical point in the parameter space where it no longer is energetically favourable for chiral symmetry breaking to occur. Near this critical point are the strongest transitions and we choose benchmarks in this region of parameter space. To ascertain the lowest temperature a confinement transition can occur we need to enforce a number of conditions
\begin{itemize}
    \item[1] To drain the entropy in the hidden sector we require kinetic equilibrium to be maintained until the freeze out temperature - typically one twentieth the pion mass \cite{Hochberg:2014dra,Hochberg:2015vrg}. 
    \item[2] We enforce the chiral limit, this means the pion mass is much lower than the critical temperature. We do not have a precise criteria, but for our present purposes require that the critical temperature to be ten times the pion mass. 
    \item[3] Freezeout needs to occur well before BBN. To ensure safety we enforce a conservative bound that $T_{\rm FO}>2$ MeV. 
\end{itemize}
Let us begin with the first criteria. We must ensure that kinetic equilibrium is maintained at the freeze-out temperature to ensure the entropy is drained . We consider dark photon mediation between mesons and SM fermions as the channel to maintain equilibrium \cite{Hochberg:2015vrg,Kuflik:2015isi}. This is achieved when,
\begin{equation}
    \frac{5 \zeta (5) }{4} m_\pi \Gamma _{\rm scatt} \lesssim \frac{H(T_f) m_\pi ^2}{T_f}
\end{equation}
where
\begin{equation}
     \Gamma _{\rm scatt} =\sum _f \frac{\sigma _f v_{\rm rel}}{E_f^2}\frac{12 \zeta (5) }{\pi ^ 2} \frac{15}{16} T_f^5
\end{equation}
and
\begin{eqnarray}
\sigma _f v_{\rm rel}&=&8 \pi \alpha _D \alpha \sum \frac{Q_\pi ^2}{N_\pi} \frac{1}{c _\chi ^2} \left[ - \left( \frac{c _\zeta ^2}{m_V^2}+ \frac{s_ \zeta ^2}{m_Z^2} \right) c_W t_\chi Q_f  \right. \nonumber \\  && \left. + \left( \frac{c_\zeta (s_\zeta +s_W t_\chi c_\zeta )}{m_V^2} - \frac{s_\zeta (c_\zeta -s_Wt_\chi s_\xi )}{m_Z^2} \right) \frac{1}{s_W c_W} (I_3^f-Q_f s_W^2) \right] E_f^2 \ . 
\end{eqnarray}
In the above we follow the notational conventions of ref. \cite{Hochberg:2015vrg}. Specifically, $(s_x,t_x,c_x)$ denote trigonometric functions where the subscript denotes the argument. The angle, $\chi $ is the mixing 
\begin{equation}
    L \supset -\frac{\sin \chi }{2} B^{\mu \nu } A^{\mu \nu } \ , 
\end{equation}
and 
\begin{equation}
    \tan 2 \zeta = \frac{m_Z^2 s_W \sin 2 \chi}{ m_V^2 -m_Z^2(c_\chi ^2-s_W^2s _\chi ^2)} \ . 
\end{equation}
Furthermore, $\pi$ denotes the dark pion, $\alpha _D$ denotes the dark sector coupling strength and all Standard Model notation is conventional. It is straightforward to achieve kinetic equilibrium and avoid all possible dark photon constraints \cite{Fradette:2014sza} for a remarkably low freezeout temperatures when the mixing is of the order $\chi \sim 10^{-4}$. \par 
The second and third constraints are arguably the most restricting. If the freezeout temperature is at least 2 MeV, this implies a pion mass of at least 40 MeV which in turn implies a critical temperature of 400 MeV\footnote{We hope this to be quite conservative since the pion mass is the geometric mean of the quark mass and the confinement temperature which is larger than the quark mass.}. This of course does not necessarily exclude interesting gravitational wave signatures from a SIMP with light quarks involving a much lower critical temperature. The weakness of the limit arising from kinetic equilibrium at freeze-out is very promising. However, the methods for calculating such a gravitational wave signal is even more uncertain than what we present as the chiral limit becomes invalid. With these restrictions in mind, we therefore present the conservative benchmarks presented in Table~\ref{tab:benchmarks} with the GW signal plotted in Figs. \ref{fig:benchmarks} and \ref{fig:benchmarks2}. Note that even THEIA requires a reasonably strong phase transition, so in general the percolation temperature can be well below the transition temperature. We take the chiral limit as being valid when the {\em critical} temperature is $400$ MeV and the transition occurs when kinetic equilibrium is still possible. This can mean in some benchmarks that the mass of the quarks are not too light compared to the transition temperature. There is a theoretical uncertainty that arises from this that needs to be settled either by lattice modelling or substantial theoretical progress.
\begin{table}[]
    \centering
    \begin{tabular}{c|c|c|c|c|c}
    Model &    $N_C$ & $N_F$ & $\Lambda $ & parameters  \\ \hline  
        LSM &  N/A & 4 & - & $m_A/m_\phi =51.66$  \\
        NJL ($B_2$) & 3 & 3 &  $1.24$ GeV &$G^{-1}=0.48$ GeV${}^{2}$, $G_D^{-1}=-0.042$ GeV${}^{5}$  \\
        NJL ($B_3$)& 5 & 3 &   $4$ GeV  &$G^{-1}=9.79$ GeV${}^{2}$,  $G_D^{-1}=-14.72$ GeV${}^{5}$ \\
        pNJL ($B_4$) & 3 & 3 &   $6.2$ GeV  &$G^{-1}=11.97$ GeV${}^{2}$,  $G_D^{-1}=-131.69$ GeV${}^{5}$  \\
    \end{tabular}
    \caption{Benchmarks for SIMP models with light quarks. In each case the critical temperature is set to the conservative limit of 400 MeV (see text for explanation). The number of flavours is denoted by $N_F$ and the number of colours by $N_C$. The parameters are found by tuning the critical temperature to 400 MeV and notational conventions are taken from ref. \cite{Hochberg:2015vrg}.}
    \label{tab:benchmarks}
\end{table}

\subsection{Asymmetric dark matter and solitosynthesis}
If a phase transition has tunneling rate that never becomes large enough compared to the Hubble time to percolate, the phase transition can still complete through a process known as solitosynthesis so long as there is a conserved charge to stabilize sub-critical bubbles \cite{Kusenko:1997hj}. If there is a conserved charge, the free energy of a sub-critical bubble is stabilized by a term proportional to the charge density inside the bubble - such a term diverges for vanishing radius. Such a stable field configuration is known as a Q-ball \cite{Coleman:1985ki}.  As the Universe cools, some sub-critical bubbles become critical and explode, completing the phase transition. The phase transition is typically strong and long lasting \cite{Croon:2019rqu}.

\par
\begin{figure}
    \centering
    \includegraphics[width=0.8\textwidth]{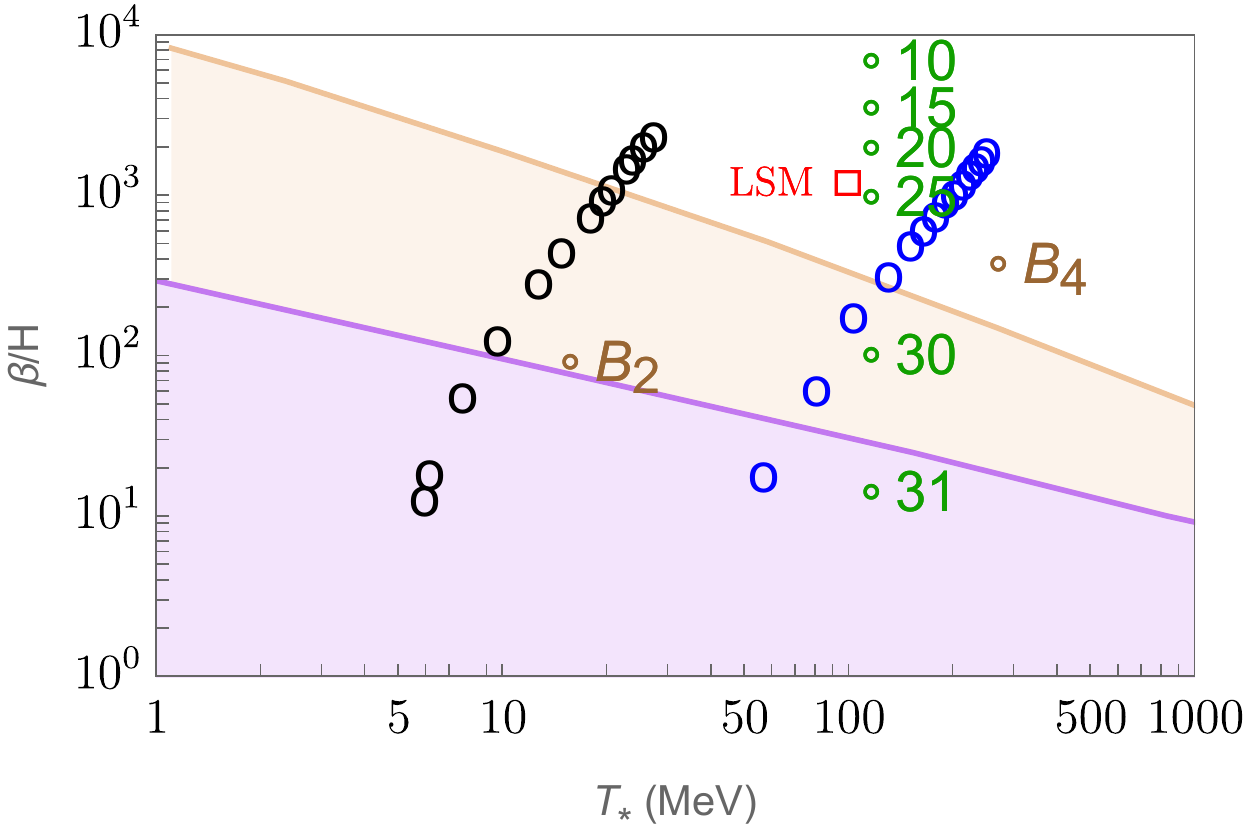}\\ \vspace{2mm}
    \caption{Sensitivity to thermal parameters with $\alpha = 1$ and benchmarks for a phase transition from solitosynthesis with asymmetry of $10^{-7}/10^{-8}$ (black/blue circles) glueballs for SU(N) (Olive numbers) and confinement in the LSM model (red squares) and the (p)NJL model. Benchmarks for the (p)NJL model are given in Table \ref{tab:benchmarks} are denoted by $B_x$.  All benchmarks are not excluded from current PTAs. More details on the models producing these transitions are given in the text. } 
    \label{fig:benchmarks}
\end{figure}

\begin{figure}
    \centering
     \includegraphics[width=0.8\textwidth]{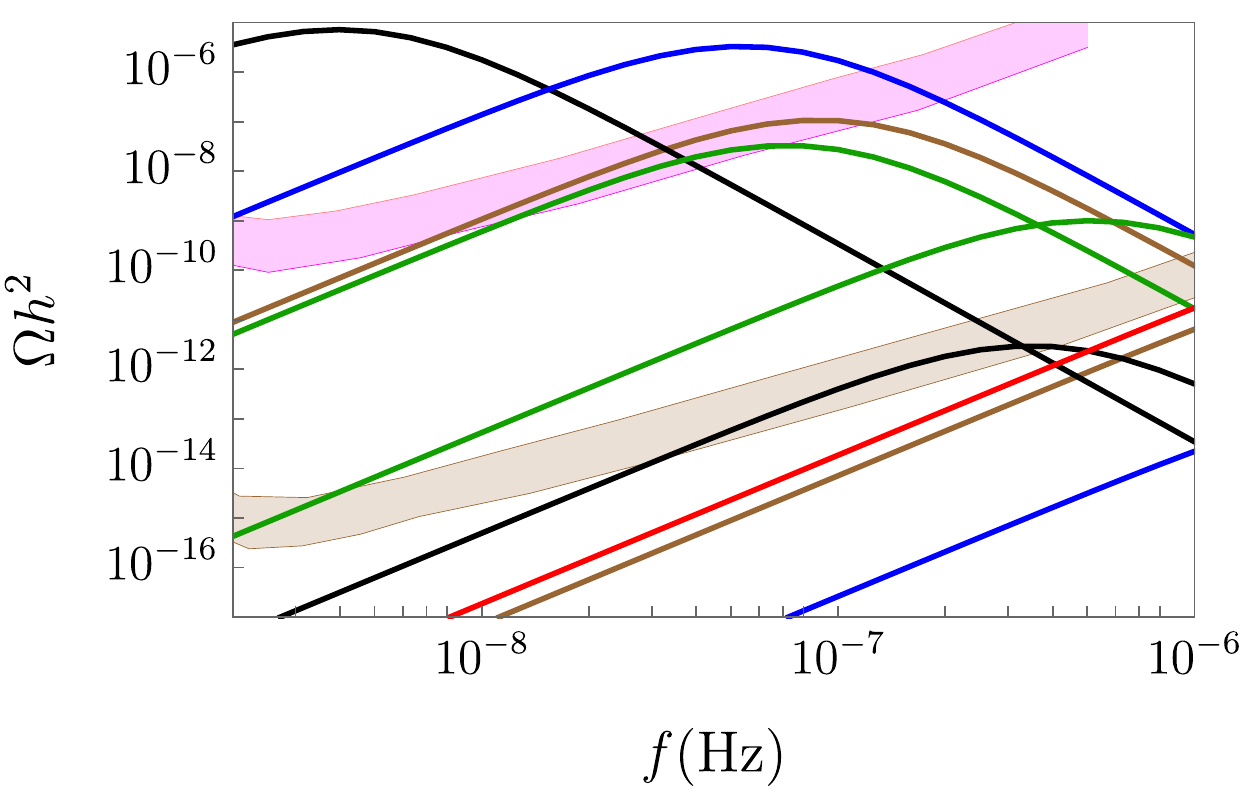}
    \caption{Same models with the same color coding as Fig. \ref{fig:benchmarks} against Gaia and THEIA sensitivity curves. In the case of Glueballs we take $N_C=30$ and $N_C=31$. For solitosynthesis we take the strongest and fifth strongest phase transitions for both values of the dark asymmetry.} 
    \label{fig:benchmarks2}
\end{figure}

  If dark matter is asymmetric, matching the dark matter to its observed value fixes a relationship between the asymmetry and the mass of the dark matter,
\begin{equation}
    Y_{\rm DM} m_{\rm DM} \sim  5\times 10^{-10} \ {\rm GeV} \ . 
\end{equation}
The transition temperature is generally within an order of magnitude of the dark matter. This makes pulsar timing arrays ideal for probing such a phase transition that has an asymmetry $Y _{\rm DM} \sim  O(10^{-7})$. Gaia is ideal for probing such asymmetric dark matter scenarios when $Y _{\rm DM} \sim O(10^{-7}-10^{-8})$. Solitosynthesis occurs in a context where the conventional tunneling rate is too slow compared to Hubble. This typically occurs when there is a tree level barrier between a true and false vacuum. The simplest possibility is a dark sector Higgs augmented by a dimension six operator\footnote{A UV completion with the same thermodynamic properties is not guaranteed, but much easier in a dark sector than a visible sector \cite{Postma:2020toi}.}
\begin{equation}
    V(h_D, \phi) = \kappa \phi ^2 h_D^2 + \Lambda ^4 \left[ (2-3 \alpha ) \left( \frac{h_D}{v_D} \right) ^2 - \left(\frac{h_D}{v_D} \right)^4+ \alpha \left(\frac{h_D}{v_D}\right)^6\right]   \ .
\end{equation}
We present the benchmarks that use the same model as Ref. \cite{Croon:2019rqu} with two benchmarks presented in Table \ref{tab:solitosynthesis} where the parameter $\alpha$ is varied between $\alpha \in (0.5,2/3)$. The sensitivity of Gaia and THEIA to these benchmarks are given in \ref{fig:benchmarks} and Fig. \ref{fig:benchmarks2}. 
\begin{table}[]
    \centering
    \begin{tabular}{c|c|c|c}
        v & $\Lambda$ & $g$ & $\eta $   \\ \hline
        0.06 GeV & 0.02 GeV & 0.1 & $10^{-7}$  \\
        0.7 GeV & 0.2 GeV & 0.1 & $10^{-8}$  
    \end{tabular}
    \caption{Benchmark parameters used for solitosynthesis following the notational conventions in ref. \cite{Croon:2019rqu}.}
    \label{tab:solitosynthesis}
\end{table}

\subsubsection{Glueballs}
Hidden sectors frequently include dark gauge symmetries which can confine. If there are no light fermions in the theory, the properties of the transition can be studied on the lattice without the ambiguities that arise from trying to model fermions. Gravitational waves arising from a pure Yang-Mills confinement transition are particularly attractive as the number of parameters in the theory, specifically the rank of the group and the confinement scale, matches the number of observables in the peak amplitude and frequency of a gravitational wave spectrum. The surface tension and the latent heat of pure Yang Mills confinement transition for a SU($N_C$) has been calculated on the lattice for $N_C \lesssim 8$ \cite{Lucini:2005vg,Datta:2010sq}
\begin{equation} 
\sigma = (0.013 N_C^3 -0.104)T_C^3 \ , \quad L=\left( 0.549+\frac{0.458}{N_C^2} \right) T_C^4 \ .
\end{equation}
Furthermore, the pressure in each phase is known and the ratio of the pressure to the Stephann-Boltzmann pressure is approximately constant as a function of $N_C$ \cite{Panero:2009tv}. We use lattice results for the pressure of an SU(3) confinement transition from ref. \cite{Agasian:2017tag,Borsanyi:2012ve}. In spite of this impressive array of lattice knowledge, the surface tension and latent heat are only calculated at the critical temperature. Similarly, the pressure below the critical temperature is only defined in the confined phase. In the early Universe, the temperature is decreasing fast enough that a fair amount of supercooling can occur by the time percolation begins. To estimate the macroscopic thermal parameters at the percolation temperature requires extrapolation of the lattice results for the supercooled phase. This can be estimated either through a toy potential for the Polyakov loop whose properties match lattice results - either a potential based on the Haar measure \cite{Huang:2020mso,Kang:2021epo} or a polynomial potential \cite{Halverson:2020xpg,Kang:2021epo}. This presents some challenges, as the growth in the surface tension with $N_C$ can be in tension with the fact that the pressure is decreasing much faster than $T^4$ near the critical temperature, combined with the fact that the absolute value of the pressure is expected to be negative for strongly super cooled transitions \cite{Cohen:2020tgr}. Another approach is to use classical nucleation theory and extrapolate the pressure to the super cooled phase. 

Since we require a fairly strong transition for it to be visible, we require a fair amount of super-cooling which unfortunately means that any method used will need to be taken with a healthy grain of salt until there are lattice calculations for supercooled transitions. Here we will review classical nucleation theory and apply it to the case of glueball nucleation. The free energy of a bubble of radius $R$ is given by
\begin{equation}
    F= -\Delta p \frac{4 \pi}{3} R^3 +\sigma 4 \pi R^2 \ .
\end{equation}
Here, $\Delta p$ is the pressure difference between the phases. There is a critical value of $R$ above which it becomes energetically favourable for the bubble to expand as the pressure overwhelms the surface tension. The nucleation rate per unit volume, per unit time is then given by the exponential of the free energy of a critical bubble divided by the temperature
\begin{equation}
\Gamma = T^4 e^{-16 \pi \sigma ^3 /(3 T \Delta p ^2) } \ .
\end{equation}
To extrapolate the pressure below the critical temperature, we use a linear extrapolation of the pressure, allowing it to go negative. A negative pressure in the supercooled deconfined phase was recently argued for in ref. \cite{Cohen:2020tgr}. Finally, the inverse time scale of the transition can then be calculated in the usual way $\beta / H = T d [F/T]/dT$. \par 

\begin{figure}
    \centering
    \includegraphics[width=0.8\textwidth]{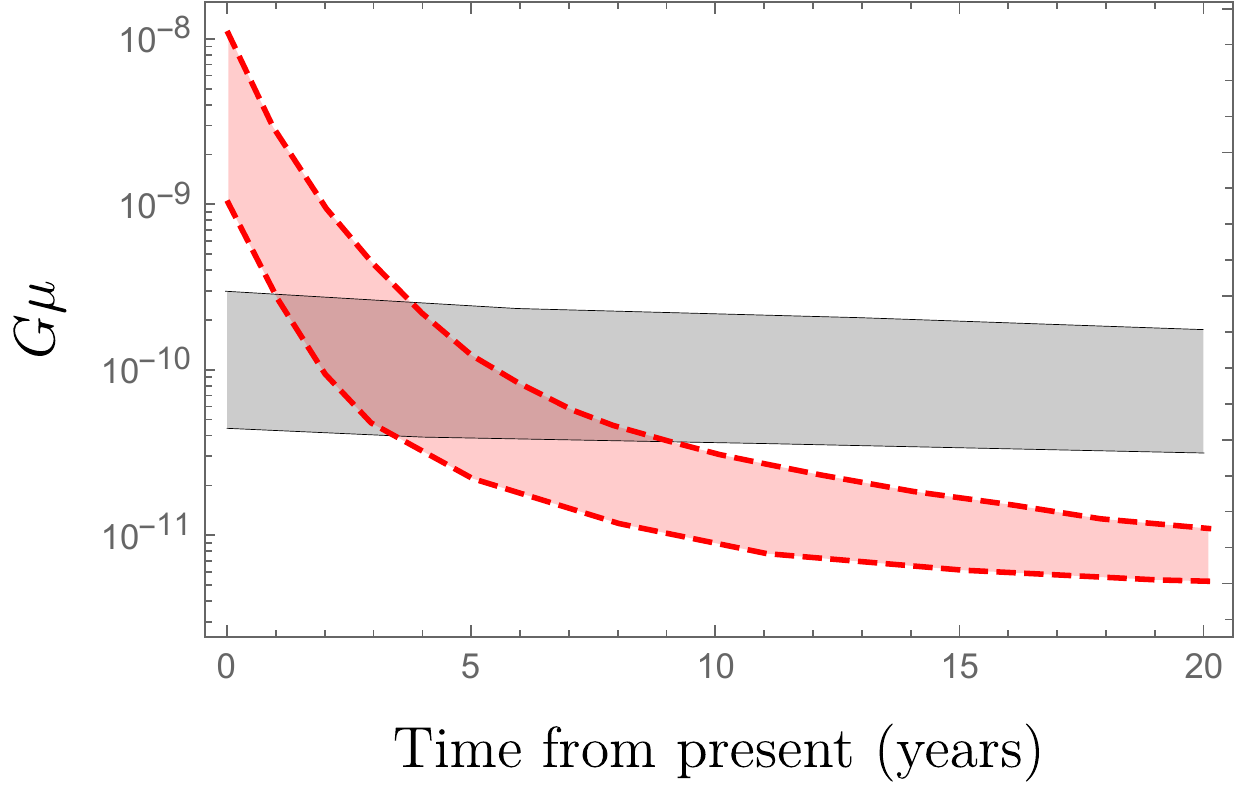} 
    \caption{ Current and projected reach of NANOGrav (gray) against projected reach of Gaia (red) for cosmic strings with the bands corresponding to 
    a signal to noise ratio SNR $\in (1,10)$.}
    \label{fig:strings}
\end{figure}

\begin{figure}
    \centering
    \includegraphics[width=0.75\textwidth]{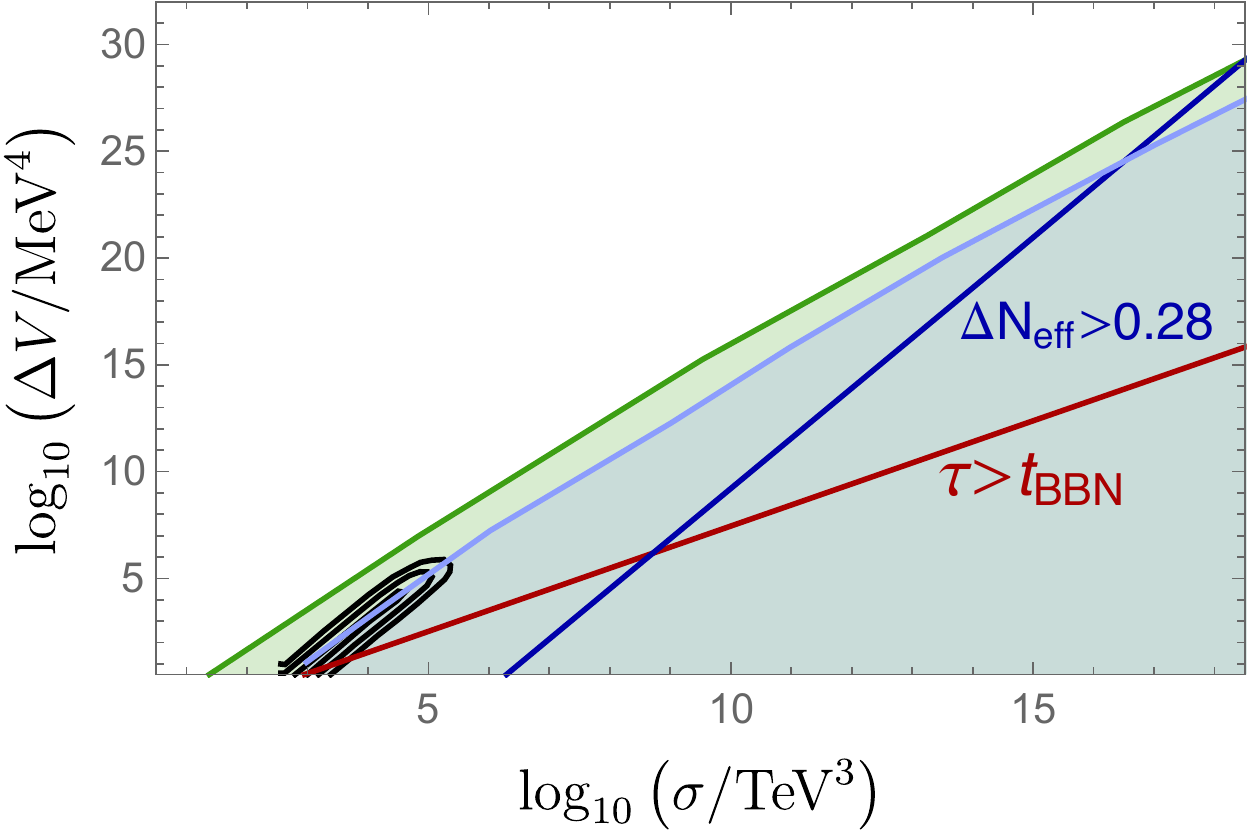}
    \caption{ Bottom: Domain walls, Black lines  give the 1,2,3 $\sigma$ fits to the potential NANOGrav signal as given in ref \cite{Bian:2020bps}, olive region is explored by THEIA, light blue region by Gaia. Below the red line the domain wall decays after BBN and below the Brown line the gravitational wave signal is too strong to be compatible with bounds on $\Delta N_{\rm eff}$}
    \label{fig:DW}
\end{figure}

While Gaia and THEIA are sensitive to phase transitions occurring at slightly higher frequencies than what would be visible to pulsar timing arrays, it is still optimal for detectability for the transition to occur as late as possible. If the glueballs are absolutely stable on cosmologically relevant lifetimes (i.e. $\tau \gtrsim 10^{25}$s to avoid cmb constraints \cite{Chen:2003gz,Fradette:2014sza,Yang:2015cva,Coffey:2020oir}), such glueballs can in principle be dark matter. However, the gravitational wave spectrum is suppressed by eight powers of the temperature ratio between the two sectors as it depends on the square of the latent heat. The temperature ratio is required to be less than one in order to avoid overclosure \cite{Forestell:2016qhc,Forestell:2017wov}. Alternatively, one can make the glueballs short lived, in which case their decay rate must be larger than Hubble before BBN in order to avoid changing the expansion rate of the Universe during BBN. If one works through a Higgs portal, the decay rate for a glueball of around the QCD scale is \cite{Juknevich:2009ji,Forestell:2016qhc}
\begin{equation}
    \Gamma \sim \frac{N^2_C \Lambda _ x^7 y^2}{\Lambda ^4 v_h ^2} \ ,
\end{equation}
where $N_C$ is the number of colors, $\Lambda _x$ is the dark confinement scale (we have made the approximation that this is equivalent to the glueball mass) and $y$ is the relevant Yukawa interactions. One can compare this rate to Hubble to find that the minimum scale is the muon threshold, that is $\Lambda \gtrsim 2 m_\mu$. In the case of vector boson mixing, the rate of Glueball decay is \cite{Juknevich:2009gg,Forestell:2016qhc}
\begin{equation}
    \Gamma \sim {\alpha _x^2 \alpha _i^2 , \alpha _x^3 \alpha _Y} N^2 \frac{\Lambda ^9_x}{\Lambda ^8}
\end{equation}
which allows for a confinement scale of around $100$ MeV for a cutoff scale of around $100 $ GeV. 

We therefore take this latter value to approximate our confinement transition temperature, though note that it is an approximation and a more detailed calculation could move the lowest possible confinement temperature somewhat.

\section{Cosmic Defects and NANOGrav}\label{sec:defects}

The recent 12.5 year updated results of NANOGrav observed a stochastic background that is at odds with expected stochastic backgrounds from super massive black hole mergers \cite{Arzoumanian:2020vkk}. A promising explanation is that the almost scale invariant background generated by cosmic strings \cite{Ellis:2020ena,Blasi:2020mfx,Datta:2020bht,Chakrabortty:2020otp,Samanta:2020cdk,King:2020hyd}. Another possibility is a metastable cosmic string \cite{Buchmuller:2020lbh} or a domain wall \cite{Bian:2020bps,Craig:2020bnv}, a strong first order phase transition \cite{Brandenburg:2021tmp,Li:2021qer,Lewicki:2020azd,Neronov:2020qrl,Nakai:2020oit,Addazi:2020zcj} or primordial black holes \cite{Kohri:2020qqd,Domenech:2020ers,Sugiyama:2020roc,DeLuca:2020agl}. In the case of metastable strings or domain walls, the low frequency Gravitational wave power spectrum obeys a cubic scaling. In this case, astrometry is an ideal compliment to PTAs as the sensitivity has a better scaling with frequency. We will here focus on stable cosmic strings and domain walls.

Let us begin with the case of a cosmic string. We use the velocity dependent one scale model of ref. \cite{Martins:1995tg,Martins:1996jp,Martins:2000cs} which models the evolution of a loop of length $\ell$ as
\begin{equation}
    \ell = \alpha _\ell t_i -\Gamma G \mu (t-t_i)
\end{equation}
where the emission rate, the string tension and the initial loop size is denoted by $\Gamma \sim 50$ \cite{Blanco-Pillado:2013qja,Blanco-Pillado:2017oxo}, $G \mu$ and $\alpha_\ell \sim 0.1$ respectively, with our choice for $\alpha _\ell$ motivated partly by recent simulations \cite{Blanco-Pillado:2013qja,Blanco-Pillado:2017oxo}. The frequency today ($t_0$)of the gravitational wave spectrum emitted by a given mode, $k$, can be related to the time of emission, $\tilde{t}$, via
\begin{equation}
    f = \frac{a(\tilde{t})}{a(t_0)} \frac{2 k }{\alpha _\ell t_i - \Gamma G \mu (\tilde{t} - t_i)} .
\end{equation}
The gravitational wave abundance is the sum of the gravitational wave spectrum for individual modes
\begin{equation}
    \Omega _{\rm GW} (f) = \sum _{k=1} ^\infty k \Gamma ^{(k)} \Omega _{\rm GW} ^{(k)} (f) \ ,
\end{equation}
where \begin{equation}
    \Gamma ^{(k)} = \frac{\Gamma k^{-4/3} }{\sum _{m=1}^\infty m^{-4/3}} \sim 13.9 k^{-4/3 } \ .
\end{equation}
The contribution to the spectrum from each mode has the form
\begin{equation}
    \Omega ^{(k)}_{\rm GW} (f) = \frac{16 \pi}{3 H_0 ^2} \frac{{\cal F} (G \mu )^2}{\alpha \ell (\alpha _\ell + \Gamma G \mu)} \frac{1}{f} \int _{t_F}^{t_0} d \tilde{t} \frac{C_{\rm eff} (t_i)}{t_i^4} \left( \frac{\alpha (\tilde{t})}{a(t_0)} \right)^5 \left( \frac{a(t_i)}{a(\tilde{t})} \right)^3 \Theta (t_i - t_F)
\end{equation}
where $t_F\sim 0$ is the time of formation, $C_{\rm eff}=5.4$ controls the number density of the loops and ${\cal F}\sim 0.1$ is an efficiency factor. 
The network can be described by a plateau, where for high frequency one essentially has a scale invariant spectrum that grows linearly with the symmetry breaking scale
\begin{equation}
    \Omega _{\rm GW} h^2\sim 8 \Omega _{\rm rad} h^2 \sqrt{\frac{G \mu}{\Gamma} } \ ,
\end{equation}
as $G \mu \sim v$ for a symmetry breaking scale $v$.
In the infrared, the spectrum curves downwards. The frequency at which the plateau occurs is quadratic in the symmetry breaking scale. This makes it more difficult for pulsar timing arrays to observe very low string tensions. 
We show the amount of time in years it will take for Gaia to have a competitive sensitivity to cosmic strings in Fig. \ref{fig:strings}. Note that Gaia does poorly in the next two years due to the frequency cutoff of the strain sensitivity being $1 /T$ where we took $T=2.5$ years at the present. Gaia quickly outperforms NANOGrav and should give complimentary data within 5 years. \par
Domain walls can form when the vacuum manifold is disconnected, for example when there is a spontaneously broken $Z_2$ symmetry. The energy density in a domain wall scales inversely with the scale factor which means that they quickly dominate the energy density of the Universe unless they are unstable. One possibility is that the symmetry was only approximate. In the case of a $Z_2$ discrete symmetry one can for example have $V(\phi _{\rm min}) = V(-\phi _{\rm min}) \pm \Delta V$. In this case the domain wall can collapse and the resulting gravitational wave power spectrum is obeys a broken power law \cite{Hiramatsu:2013qaa,Kadota:2015dza,Zhou:2020ojf} 
\begin{eqnarray}
 &&  \Omega _{\rm GW}(f) h^2 =  \nonumber \\ 
 &&5.2 \times 10^{-20} \epsilon _{\rm GW} {\cal A} ^4\left( \frac{g _\ast }{10.75} \right)^{1/3}  \left( \frac{\sigma}{ {\rm TeV}^3} \right)^{4} \left( \frac{{\rm MeV}^4}{\Delta V} \right)^2\left[ \Theta (f-f_p) \left(\frac{f}{f_p}\right)^{-1}+\Theta (fp-f) \left(\frac{f}{f_p} \right)^3 \right] \ , \nonumber \\
\end{eqnarray}
where $ \epsilon _{\rm GW} \sim 0.7$ \cite{Hiramatsu:2013qaa} is an efficiency parameter, ${\cal A}\sim 1.2$ is an area parameter, $\sigma $ is the tension in the wall and the peak frequency is given by
\begin{equation}
    f_p = 3.99\, {\rm nHz} \, A^{-1/2} \left( \frac{\rm TeV^3}{\sigma } \right)^{1/2} \left( \frac{\Delta V}{\rm MeV^4} \right)^{1/2} \ .
\end{equation}
Note that it is possible for the gravitational wave abundance to be an unacceptably large amount of radiation, changing the expansion rate of the Universe at recombination. The constraint on the number of relativistic degrees of freedom can be recast into a constraint on the gravitational wave power spectrum \cite{Opferkuch:2019zbd}
\begin{equation}
    \Delta N_{\rm eff} = \frac{8}{7} \left(  \frac{11}{4}\right) ^{4/3} \int \frac{df}{f} \frac{\Omega _{\rm GW}}{\Omega _{\rm rad}} \ . 
\end{equation}
In Fig. \ref{fig:DW} we show the projected sensitivity of Gaia and THEIA as well as the current best fit to the NANOGrav signal from ref. \cite{Zhou:2020ojf} and the contraint from $\Delta N_{\rm eff}$.
\section{Primordial black holes}\label{sec:pbh}

The extra components beyond the Standard Model responsible for the strong phase transitions that give rise to a significant Stochastic Gravitational Wave Background (SGWB), as discussed in the previous sections, also modify the number of relativistic degrees of freedom and rate of expansion of the universe around the phase transition. The contribution of these new degrees of freedom change the total pressure and energy density of the plasma and thus the equation of state of the universe. We will assume that the extra phase transition responsible for the SGWB occurs around that of the QCD quark-hadron transition, so that the effect is enhanced. Note that the QCD horizon size gives both a solar mass scale ($M_{\rm hor} = (c^3/G)\,t_{\rm QCD} \sim 1\,\Msun$) and nanohertz frequencies ($f_{\rm peak} = 1/(2t_{\rm QCD}(1+z_{\rm QCD}))) \sim 10$ nHz), which connects LIGO/Virgo GWTC-2 black holes with the Pulsar Timing Array and Gaia/THEIA surveys.

\begin{figure*}
    \centering
    \includegraphics[width=0.48\textwidth]{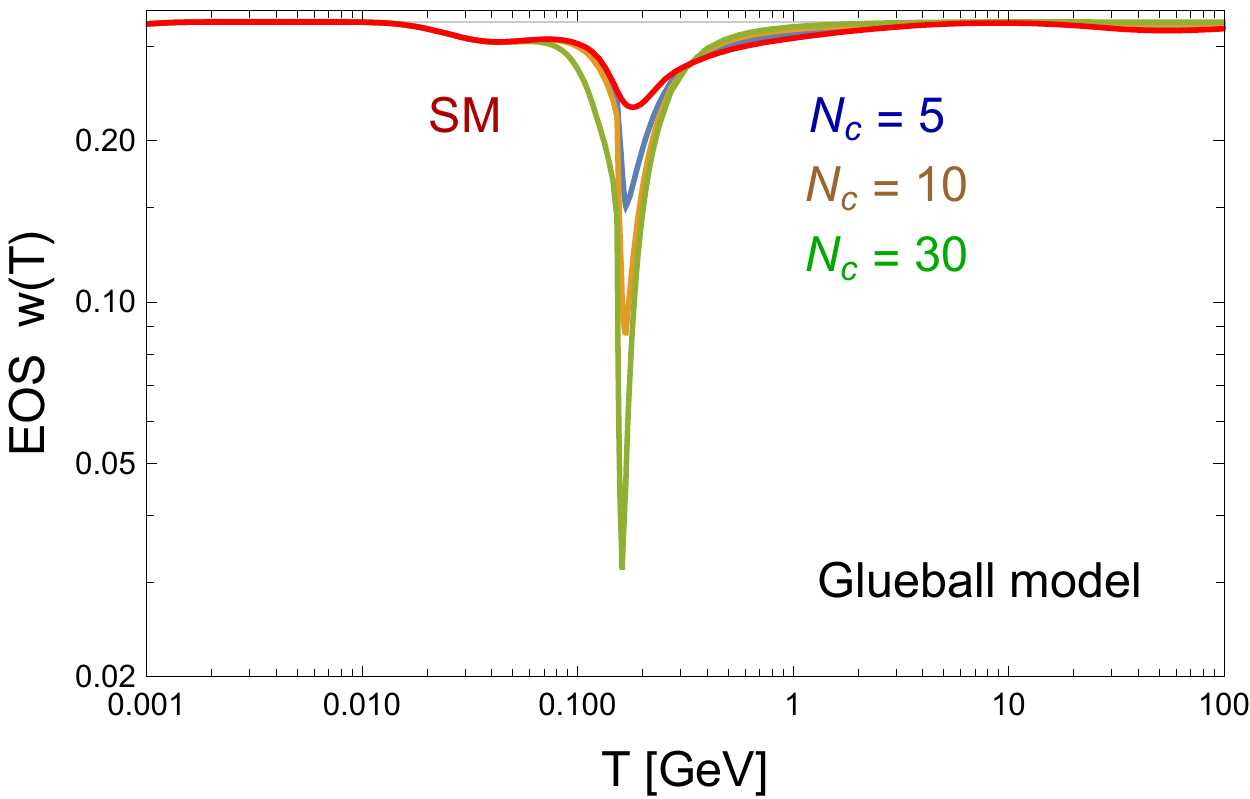}
    \includegraphics[width=0.48\textwidth]{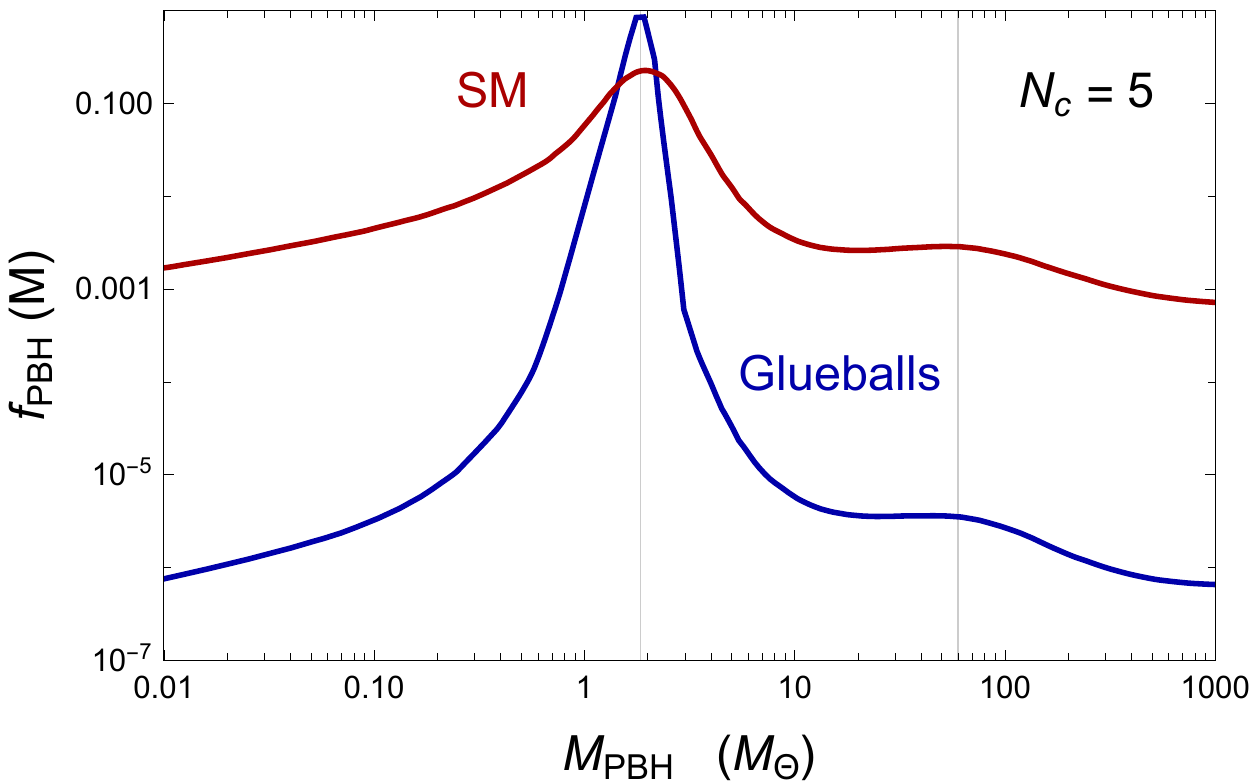}\\[5mm]
    \includegraphics[width=0.48\textwidth]{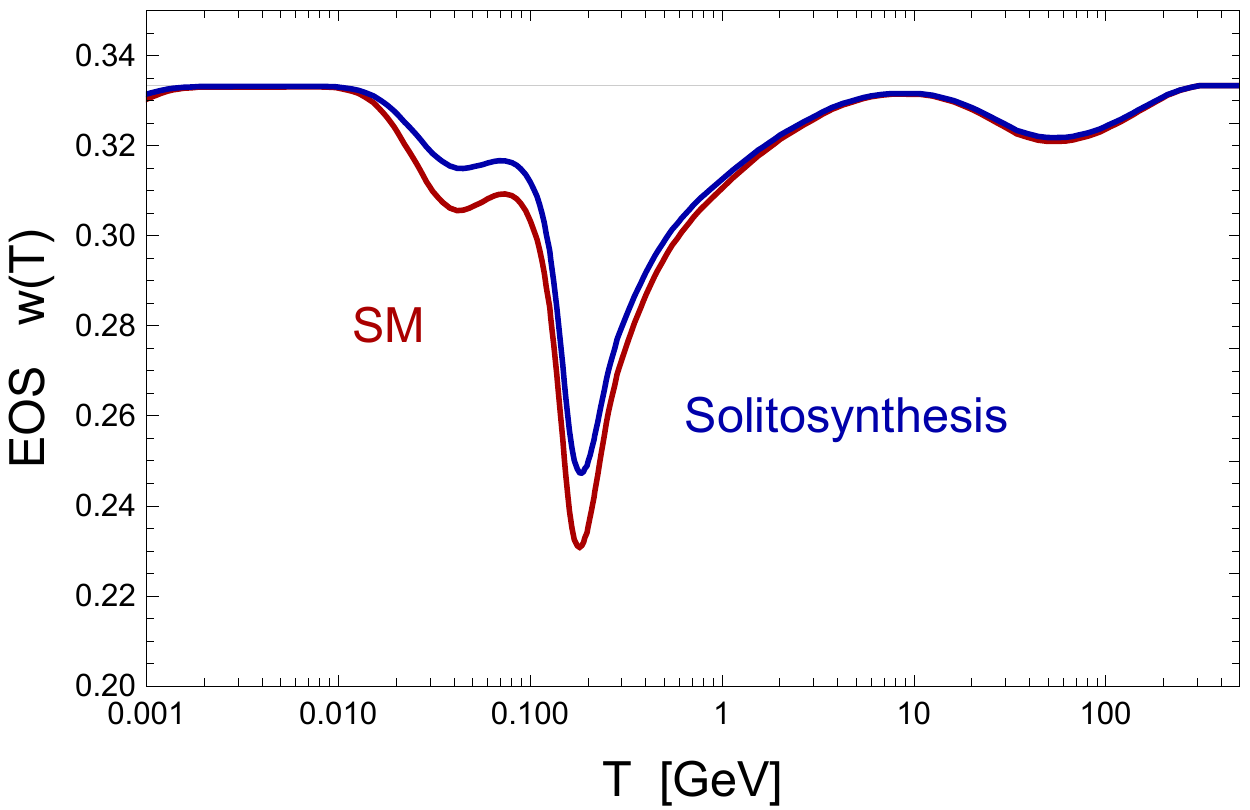}
    \includegraphics[width=0.48\textwidth]{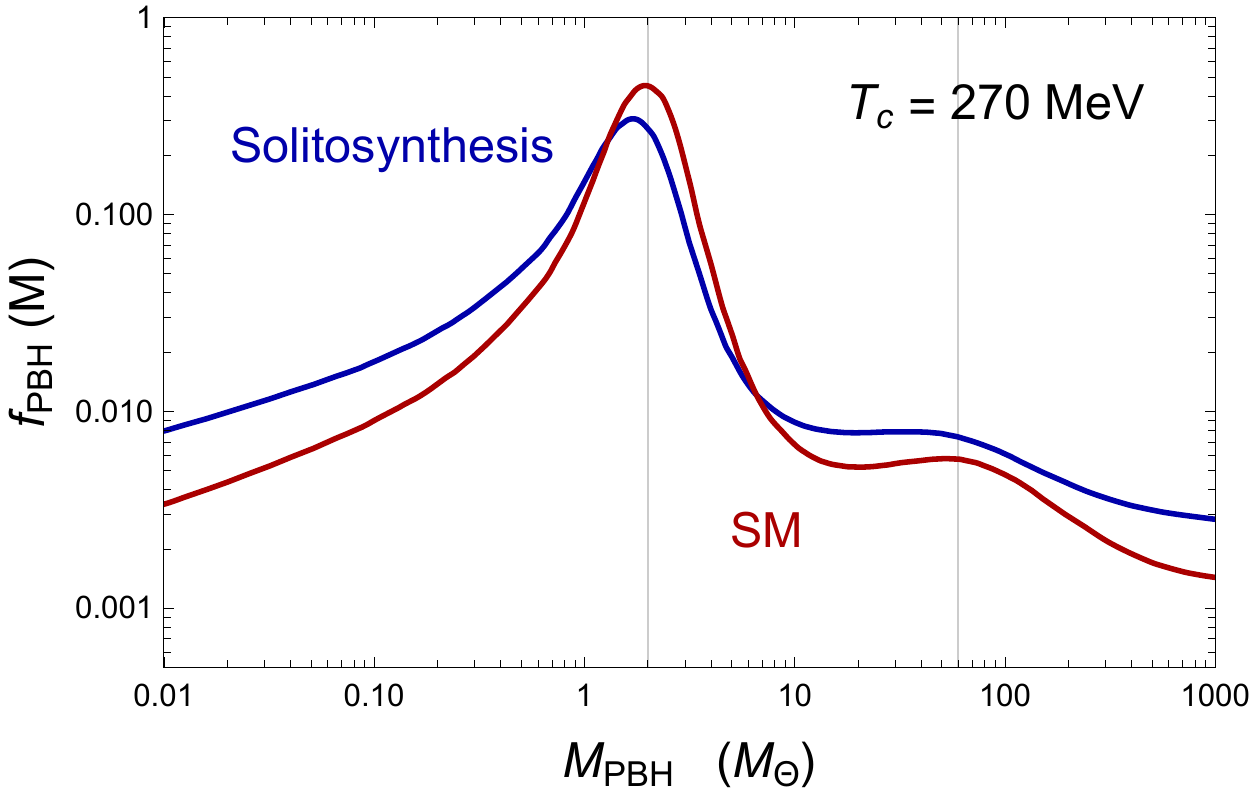}\\[5mm]
    \includegraphics[width=0.48\textwidth]{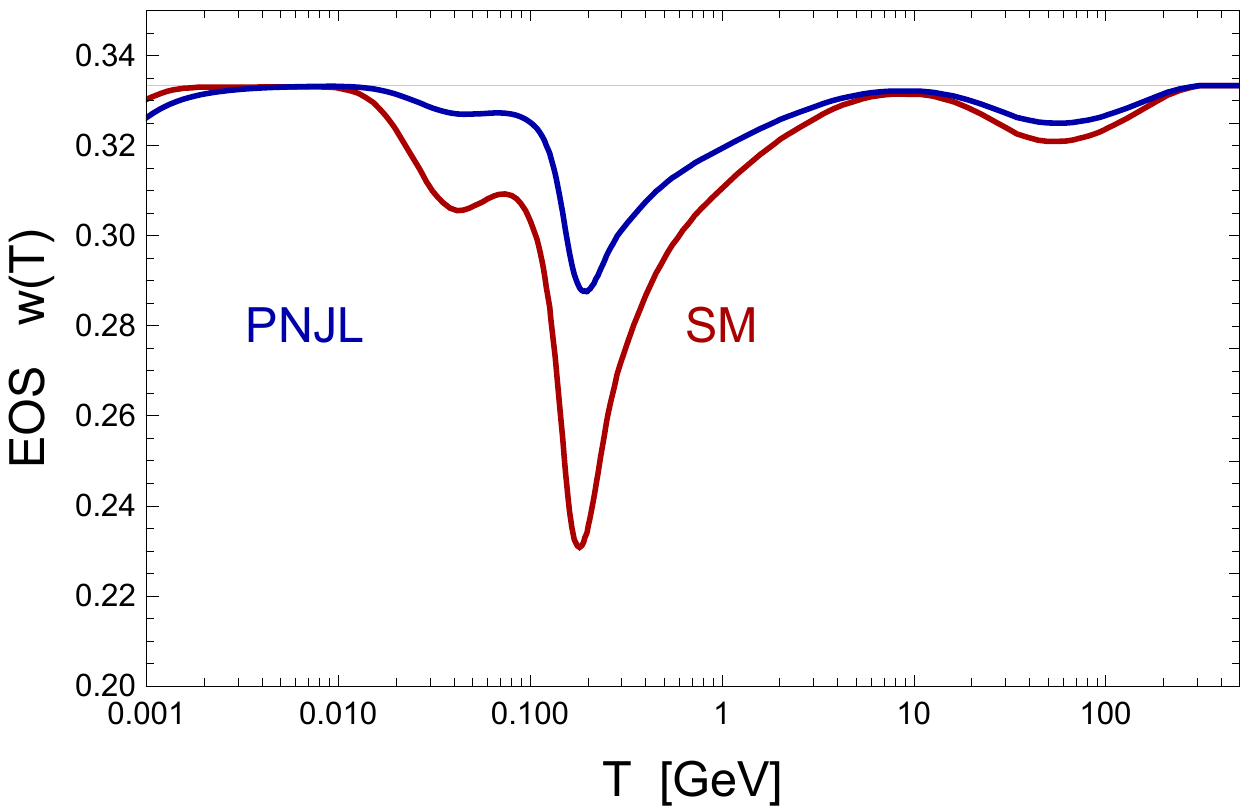}
    \includegraphics[width=0.48\textwidth]{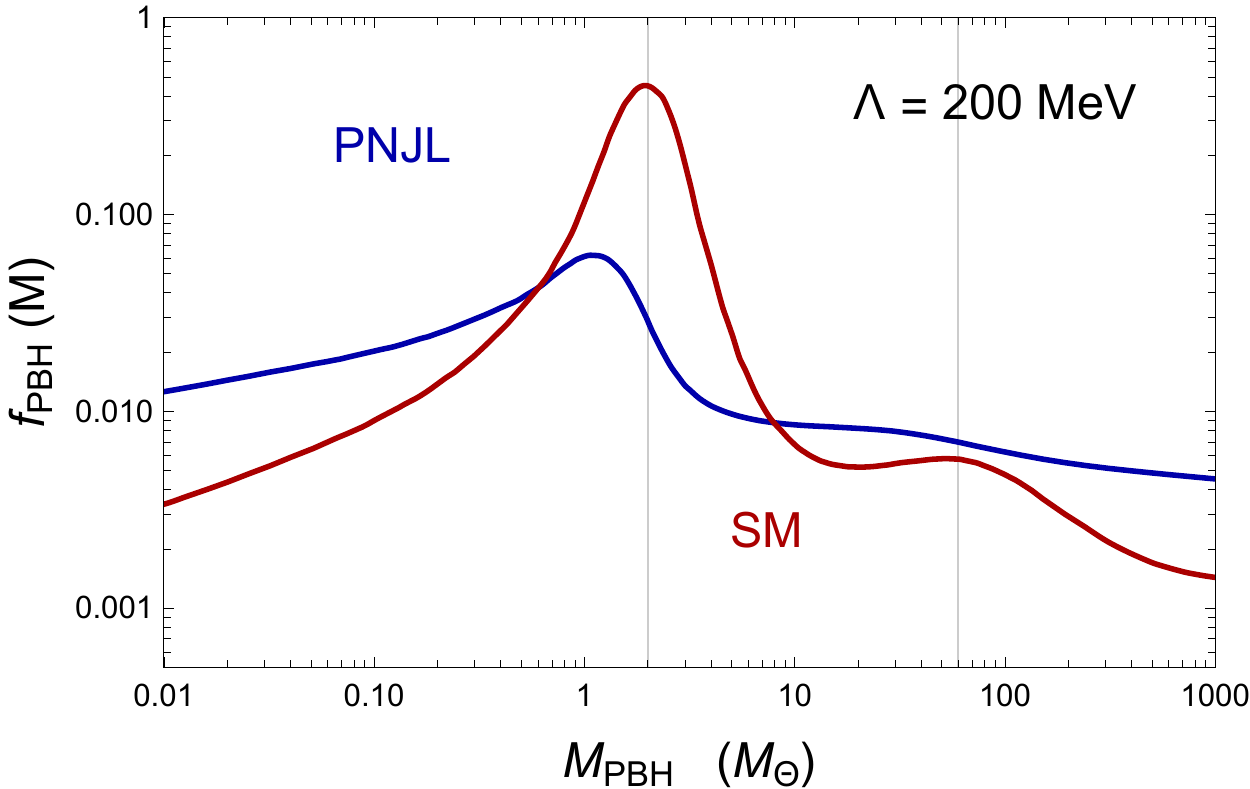}\\[5mm]
    \caption{The equation of state around the phase transition, as a function of the temperature of the Universe, and the PBH mass fraction distribution as a function of BH mass in solar units.}
    \label{fig:PBH}
\end{figure*}

A rapid change in the equation of state of the plasma during the QCD transition will decrease the radiation pressure and thus will allow for a few (horizon) domains to collapse to form primordial black holes~\cite{Carr:2019kxo,Garcia-Bellido:2019tvz,Bodeker:2020stj}. The sensitivity to the equation of state is exponential and thus a small change in radiation pressure can completely change the PBH mass spectrum. The threshold for gravitational collapse $\delta_c ( w )$ is a function of the equation-of-state parameter $w( T )$, see Ref.~\cite{Musco:2012au}, so the thermal history of the Universe can induce pronounced features in the PBH mass function even for a uniform power spectrum of fluctuations. The reason being that, if the PBH form from Gaussian inhomogeneities with root-mean-square amplitude $\delta_{\rm rms}$, then the fraction of horizon patches undergoing collapse to PBH when the temperature of the Universe is $T$ should be \cite{Carr:2019kxo}
\be
\beta( M )\approx {\rm Erfc}\!
\left[\frac{\delta_c\big( w[ T( M ) ] \big)}{ \sqrt{2} \, \delta_{\rm rms}( M )}\right]\,,
\label{eq:beta}
\ee
where the temperature is related to the PBH mass by $T\approx 200\,\sqrt{\Msun / M\,}\;{\rm MeV}$.
This shows that $\beta( M )$ is exponentially sensitive to $w( M )$. The present CDM fraction for PBHs of mass $M$ is then
\be
\fPBH(M) \equiv \frac{1}{\rho_{\rm CDM}}\frac{d\,\rho_{\rm PBH}(M)}{d\ln M }
\approx \frac{2\OM}{\OCDM}\,\beta(M)
\sqrt{\frac{M_{\rm eq}}{M}}\,,
\label{eq:fPBH}
\ee
where $M_{\rm eq} = 2.8 \times 10^{17}\,\Msun$ is the horizon mass at matter-radiation equality and $\rho_{\rm CDM}$ is the CDM density. 
We have computed the change in EOS,
\begin{equation}
w_i(T) = \frac{p_{\rm SM}(T) + p_i(T)}{\rho_{\rm SM}(T)+\rho_i(T)}\,,
\end{equation}
during the thermal evolution of the universe in the three cases discussed above: Glueballs, Solito\-synthesis and the PNJL model. Each one gives a different mass spectrum for PBH, when taking into account the exponentially sensitive collapse to PBH as the radiation pressure varies accross the corresponding transitions. In the case of Glueballs, it generates for large $N_c$ a delta function of PBH at a particular scale, which is strongly constrained by astrophysical and cosmological observations. On the other hand, the Solitosynthesis model changes only very slightly the PBH mass spectrum from that of the SM. Finally, the PNJL model is sufficiently different from the SM that the search for signatures of differences in the mass spectrum will eventually be detected in the mass distribution of black holes from GW interferometers like LIGO/Virgo/KAGRA, as well as with current microlensing surveys like OGLE/GAIA and future surveys like LSST.

To test the robustness of our PBH results, we changed the value of the critical temperature $T_c$ and the scale $\Lambda$ to lower values (but still above the BBN scale), in order to generate peaks at higher masses, and we found no significant differences with the curves shown in Fig.~6, which makes these mass spectra rather robust.

\section{Conclusions}\label{sec:conclusions}

In this paper, we have discussed the potential of a stochastic gravitational wave background detection at Gaia or future THEIA mission to probe early universe physics. These astrometric surveys open up a new frequency window of gravitational wave detection between LISA and pulsar timing arrays. We pointed out that there are many scenarios of phase transitions associated with dark matter such as SIMP or asymmetric dark matter, solitosynthesis, cosmic defects, and primordial black holes that can lead to detectable signals. Uncertainties in theoretical predictions are spelled out. We hope our work stimulates further discussions for the design of the missions and survey strategy in order to capture exciting physics potential.

\appendix{}
\section{Low energy effective models for modeling confinement transitions}

\subsection{Linear sigma model}
For convenience we give just the details of the potential for four light flavours
\begin{equation}
    V(\phi , T) = V_0(\phi) + V_T (\phi , T) 
\end{equation}
where
\begin{eqnarray}
V_0(\phi ) &=& \frac{1}{32} \left( - 16 m _\sigma ^2 \phi ^2 + (\kappa + 4 \lambda - \mu _\sigma )\phi ^4  \right) \\ 
V_T(\phi , T) &=& \bar{J}_B(m_\phi^2 /T^2) + \bar{J}_B(m_\eta^2 /T^2) + 15 \bar{J}_B (m_{X_8}^2/T^2) +8 \bar{J}_B(m_{X_3}^2/T^2)+15 \bar{J} _B (m_{\pi _8}^2 /T^2) \nonumber \\ 
&& +8 \bar{J} _B ( m_{\pi _3}^2 /T^2 ) + \bar{J}_B( m_{\eta _\psi }) + \bar{J} _B (m_{\eta _\chi} ^2 / T^2) .
\end{eqnarray}
In the above
\begin{equation}
    \bar{J} _B (z^2) = \frac{T^4}{2 \pi ^2} \int  _0 ^\infty dx x^2 \log \left[ 1- e^{\sqrt{x^2 + z^2}} \right] - \frac{T}{12 \pi} \left( (z^2 + \frac{1}{12}(3 \kappa + 17 \lambda) T^2 )^{3/2})- z^3\right) 
\end{equation}
and the masses are
\begin{eqnarray}
m_\phi &=& \frac{1}{24} \left( 9 \kappa \phi ^2 + 36 \lambda \phi ^2 - 9 \mu \phi ^2 + 6 \mu _s -24 m _\Sigma  \right) \\ 
m _ \eta &=& \frac{1}{8} \left( \phi ^2 [\kappa + 4 \lambda + 3 \mu] + 2 \mu \sigma \right) - m_\Sigma  \\ 
m_{X_8} &=& \frac{1}{8} \phi ^2 \left( 3 \kappa + 4 \lambda + \mu  \right) - m _\Sigma  \\
m_{X_3} &=& \frac{1}{24} \left( 9 \kappa \phi ^2 +12 \lambda \phi ^2 + 3 \mu \phi ^2 -24 m_\Sigma  \right)  \\ 
m_{\pi _8} &=& \frac{1}{8} \phi ^2 \left( \kappa + 4 \lambda - \mu \right) - m _\Sigma \\ 
m_{\pi _3} &=& \frac{1}{24} \left(3 \kappa \phi ^2 + 12 \lambda \phi ^2 - 3 \mu \phi ^2 - 24 \mu. _\Sigma   \right) \\ 
m_{\eta _\psi} &=& \frac{1}{24} \left( 3 \phi ^2 [3 \kappa + 4 \lambda + \mu] + 18 \mu _s -24 \mu _\Sigma  \right) \\ 
m_{\eta _\chi } &=& \frac{1}{24} \left( 3 \phi ^2 [\kappa + 4 \lambda - \mu] + 18 \mu_S - 24 m_\Sigma \right)
\end{eqnarray}
\subsection{(p)NJL model}
The effective potential for the composite field, $\sigma $, in both the NJL and PNJL model is as follows
\begin{equation}
    V_{\rm eff}^{\rm (P)NJL} (\bar{\sigma } ,L , T) = V_0 ^{\rm (P)NJL} (\bar{\sigma })+V_{\rm CW} ^{\rm (P)NJL} ( \bar{\sigma }) + V_{\rm FT} ^{\rm (P)NJL} ( \bar{\sigma } , L , T)
\end{equation}
where
\begin{eqnarray}
 V_0 ^{\rm (P)NJL} (\bar{\sigma } ) &=& \frac{3}{8 G} \bar{ \sigma }^2 - \frac{G_D}{16 G^3}\bar{ \sigma }^3 \\
 V_{\rm CW}^{\rm (P)NJL} (\bar{\sigma}) &=& -\frac{3 N_C}{16 \pi ^2} \left[ \Lambda ^4 \log \left( 1+ \frac{M^2}{\Lambda ^2} \right) - M^4 \log \left( 1 + \frac{\Lambda ^2}{M^2} \right) + \Lambda ^2 M^2  \right] \\ 
 V_{\rm FT}^{\rm (P)NJL} &=& - \frac{6 T^2}{\pi ^2} \int _0 ^\infty dx x^2 \log \left( 1 + e^{-3 \sqrt{x^2 +r^2}} + 3 L e^{-\sqrt{x^2 +r^2}}+ 3 L e^{-2 \sqrt{x^2 +r^2}} \right)  \nonumber \\ 
&& + T^4 \left( - \frac{1}{2} a(T) L^2 + b(T)\log[1-6 L^2 - 3L^4 + 8 L^3] \right) . \end{eqnarray}
The last term is the potential for the glueballs and exists only in the pNJL model where the log term is from the Haar measure. We follow the strategy of ref \cite{Helmboldt:2019pan} in minimizing the potential with respect to $L$, for each value of $\bar{\sigma}$ to reduce the problem to a single field problem. One can choose coefficients such that the potential reproduces the surface tension and the evolution of the pressure for $N_C=3$
\begin{eqnarray}
 a(T) &=& a_0 + a_1 \frac{T_{C}}{T} + a_2 \left( \frac{T_{C}}{T}  \right)^2 \\ 
 b(T)&=& b_3 \left( \frac{T_C}{T} \right) ^{3} \ ,
\end{eqnarray}
where
\begin{equation}
    a_0 =3.51, \quad a_1 =-2.47, \quad a_2 = 15.2 , \quad b_3 = - 1.75.
\end{equation}
In the above $T_C$ is an input parameter that needs to match the critical temperature of the potential. For each value of $G$ and $G_D$, one can iteratively modify $T_C$ until it matches the physical critical temperature. Finally the field dependent mass is
\begin{equation}
    M = \bar{\sigma } -\frac{G_D}{8 G^2} \bar{\sigma }^2 .
\end{equation}

\par 

The field $\bar{\sigma }$ is composite, thus the Euclidean action has a non-canonical kinetic term
\begin{equation}
    S_3 [\bar{\sigma}] = 4 \pi \int r^2 dr \left[ \frac{Z_{\sigma} ^{-1}}{2} \left( \frac{d \bar{\sigma }}{ dr} \right)^2 + V_{\rm eff} (\bar{\sigma }) \right]  
\end{equation}
where
\begin{equation}
    Z_\sigma ^{-1} = - 3 N_C \left( 1 - \frac{G_D}{4 G^2} \sigma \right)^2 \left[ -2 A_0 +2 B_0 +8 C_0-2\ell _A(r) + 2 \ell_B(r)+8 \ell_C(r) \right] \ .
\end{equation}
In the above $r=|M(\bar{\sigma } )|/T$ with
\begin{eqnarray}
 A_0 &=& \frac{1}{16 \pi ^2} \left[ \log \left( 1+ \frac{\Lambda ^2}{ M^2} \right) - \frac{\Lambda ^2}{\Lambda ^2 + M^2} \right] \\ 
 B_0 &=& - \frac{1}{ 32 \pi ^2 } \frac{\Lambda ^4}{(M^2 + \Lambda ^2)^2} \\ 
 C_0 &=& \frac{1}{96 \pi ^2} \frac{3 M^2 \Lambda ^4 + \Lambda ^6}{(M^2 + \Lambda ^2)^3}
\end{eqnarray}
finally the relevant thermal integrals are
\begin{eqnarray}
\ell _A (r) &=& - \frac{1}{4 \pi ^2} \int _0 ^\infty \left( \frac{x^2}{\left( x^2+r^2 \right)^{3/2}} \frac{1}{1+{\rm exp}[\sqrt{x^2+r^2}] } \right. \nonumber \\
&& + \frac{1}{2} \frac{x^2}{x^2+r^2} \frac{1}{1+ \cosh [\sqrt{x^2 + r^2}]} \\ 
\ell _B (r) &=& \frac{r^2}{16 \pi ^2} \int _0 ^\infty dx \left( \frac{3 x^2}{(x^2+r^2)^{5/2}}\frac{1}{1+\exp [\sqrt{x^2+r^2}]} + \frac{3 x^2}{2(x^2+r^2)^2}\frac{1}{1+\cosh \sqrt{[x^2+r^2]}} \right.\nonumber \\ &&
\left. + \frac{x^2}{2 (x^2 +r^2)^{5/2}} \frac{1}{1+\cosh [\sqrt{x^2+r^2}]}  \right) \\
\ell _C (r) &=& -\frac{r^4}{96 pi ^2} \int _0 ^\infty dx \left( \frac{15 x^2}{2(x^2 + r^2)^{7/2}} \frac{1}{1+\exp [\sqrt{x^2+r^2}]} + \frac{15 x^2}{2 (x^2 + r^2)^3} \frac{1}{1+\cosh [\sqrt{x^2+r^2}]} \right. \nonumber \\ 
&& \left. \frac{3 x^2}{(x^2+r^2)^{5/2}} \frac{\tanh (\frac{1}{2}\sqrt{r^2 +x^2})}{1+\cosh [\sqrt{x^2+r^2}]} + \frac{x^2}{2 (x^2 +r^2)^2} \frac{1}{1+ \cosh [\sqrt{x^2+r^2}]} \right. \nonumber \\
&&- \frac{3 x^2}{2 (x^2 + r^2)^2} \frac{1}{(1+\cosh \sqrt{x^2 + r^2})^2}
\end{eqnarray}

\section*{Acknowledgements}

The authors would like to thank Eleanor Hall for collaboration at the early stage of this work. GW thanks David Morrissey, Oleg Popov and Djuna Croon for discussions about glueballs. JGB thanks Deyan Mihaylov for enlightening discussions on Gaia sensitivity to a SGWB and also acknowledges funding from the Research Project PGC2018-094773-B-C32 (MINECO-FEDER) and the Centro de Excelencia Severo Ochoa Program SEV-2016-0597. HM was supported by the Director, Office of Science, Office of
High Energy Physics of the U.S. Department of Energy under the
Contract No. DE-AC02-05CH11231, by the NSF grant
PHY-1915314, by the JSPS Grant-in-Aid for
Scientific Research JP20K03942, MEXT Grant-in-Aid for Transformative Research Areas (A)
JP20H05850, JP20A203, by WPI, MEXT, Japan, and Hamamatsu Photonics, K.K.

\bibliographystyle{utphys}
\bibliography{mm}
\end{document}